\documentclass[aps,prc,amsfonts,preprintnumbers,superscriptaddress,showpacs,nofootinbib]{revtex4}
 
\usepackage{graphics}
\usepackage{psfrag}
\usepackage{epsfig}
\usepackage{epsf}
\usepackage{float}

\newcommand{\beq}{\begin{equation}}
\newcommand{\eeq}{\end{equation}}
\newcommand{\beqa}{\begin{eqnarray}}
\newcommand{\eeqa}{\end{eqnarray}}
\newcommand{\nn}{\nonumber\\}

\newcommand{\tr}{\mbox{tr}}

\newcommand{\newop}[2]{\def#1{\mathop{\mathrm{#2}}\nolimits}}
\newop{\artanh}{artanh}
\newop{\det}{det}
\newop{\tr}{tr}
\newop{\diag}{diag}
\newop{\Re}{Re}
\newop{\Im}{Im}

\def\sl#1{\hspace{0.1 truecm}\slash{\hspace{-0.25 truecm}#1}}
\def\slsm#1{\slash{\hspace{-0.2 truecm}#1}}

\begin{document}

\preprint{FZJ-IKP-TH-2009-14}
\preprint{HISKP-TH-09/19}

\title{Redundancy of the off-shell parameters in chiral effective field theory\\ with explicit
      spin-3/2 degrees of freedom}

\author{H.~Krebs}
\email[]{Email: hkrebs@itkp.uni-bonn.de}
\affiliation{Helmholtz-Institut f\"ur Strahlen- und Kernphysik (Theorie)
and Bethe Center for Theoretical Physics,
 Universit\"at Bonn, D-53115 Bonn, Germany}
\author{E.~Epelbaum}\email{e.epelbaum@fz-juelich.de}
\affiliation{Forschungszentrum J\"ulich, Institut f\"ur Kernphysik (IKP-3) and
J\"ulich Center for Hadron Physics, \\
             D-52425 J\"ulich, Germany}
\affiliation{Helmholtz-Institut f\"ur Strahlen- und Kernphysik (Theorie)
and Bethe Center for Theoretical Physics,
 Universit\"at Bonn, D-53115 Bonn, Germany}
\author{Ulf-G. Mei{\ss}ner}\email{meissner@itkp.uni-bonn.de}
\affiliation{Helmholtz-Institut f\"ur Strahlen- und Kernphysik (Theorie)
and Bethe Center for Theoretical Physics,
 Universit\"at Bonn, D-53115 Bonn, Germany}
\affiliation{Forschungszentrum J\"ulich, Institut f\"ur Kernphysik (IKP-3),
Institite for Advanced Simulation (IAS-4)\\ and
J\"ulich Center for Hadron Physics, 
             D-52425 J\"ulich, Germany}
\date{\today}

\begin{abstract}
In this note we prove to all orders in the small scale expansion that all
off-shell parameters which appear in the chiral effective Lagrangian with explicit 
$\Delta(1232)$ isobar degrees of freedom can be absorbed into redefinitions of
certain low-energy constants and are therefore redundant.
\end{abstract}


\maketitle


\section{Introduction and Summary}
\def\theequation{\arabic{section}.\arabic{equation}}
\setcounter{equation}{0}
\label{sec:intro}

Chiral effective field theory (EFT) with explicit $\Delta (1232)$ isobar
degrees of freedom is usually formulated utilizing the Rarita-Schwinger
formalism (for pioneering works, see \cite{Jenkins:1991es,Hemmert:1997ye}).
In this framework, the spin-$3/2$ field is represented by a vector-spinor 
$\psi_\mu$ (here and in what follows, we omit the spinor indices). It is well known 
how to write down the most general free Lagrangian for $\psi_\mu$ which
describes the proper  
number of degrees of freedom \cite{Moldauer:1955}. The unphysical spin-$1/2$
degrees of freedom are projected out in the resulting free equations of
motion. It is considerably more difficult to ensure the decoupling of 
the unphysical degrees of freedom in the case of interacting  spin-$3/2$
fields. An elegant way to achieve this goal is to require that all
interactions have the same type of gauge invariance as the kinetic term of the
spin-$3/2$ field \cite{Pascalutsa:1998pw}. For a different strategy based on
analyzing the consistency conditions within the canonical Hamilton formalism
see Ref.~\cite{Wies:2006rv}. This requirement of gauge invariance 
is, however, not compatible with the non-linear realization of the chiral symmetry by 
Coleman, Callan, Wess and Zumino \cite{Coleman:1969sm,Callan:1969sn}, which is
commonly adopted in chiral effective field theories and ensures the chiral
invariance of the effective Lagrangian on a term-by-term basis.  
Recently we have shown that any bilinear coupling of a massive spin-3/2 field
can be brought into a gauge invariant form suggested by Pascalutsa by means of
a non-linear field redefinition \cite{Krebs:2008zb}. A similar statement can also be made for
linear couplings of the $\Delta$ to a nucleon field \cite{Pascalutsa:2000kd}. The
equivalence of the two 
formulations by means of the non-linear field redefinition implies
that S-matrix elements can be calculated from the standard effective Lagrangian
with the chiral symmetry being realized on a term-by-term basis using 
\emph{naive} Feynman rules (i.~e.~ignoring the field-dependent determinants in
the path-integral measure which may arise in the constrained quantization procedure). 

An important question which still needs to be addressed in this context is
related to the role of the so-called off-shell parameters which accompany
every interaction term in the most general effective chiral Lagrangian with
explicit $\Delta (1232)$ isobar fields, see section~\ref{sec1} for details. 
Here, an important observation was made by Ellis and Tang
\cite{Tang:1996sq,Ellis:1996bd} who showed that all three 
off-shell parameters appearing in the leading $\pi N \Delta$ and $\pi \Delta
\Delta$ Lagrangians are redundant as they can be absorbed into other
parameters in the Lagrangian, see also \cite{Pascalutsa:2000kd}. These
findings were confirmed in the explicit calculation by Fettes and 
Mei{\ss}ner \cite{Fettes:2000bb} based on the so-called small scale expansion 
(SSE) \cite{Hemmert:1997ye}. To the best of our knowledge, no rigorous proof of the
redundancy of the off-shell parameters in chiral EFT to all orders in the SSE
for arbitrary processes has been given in the literature. In this
work  we fill this gap and demonstrate explicitly that all terms in the
effective chiral Lagrangian which give rise to the explicit dependence on the
off-shell parameters can be eliminated by means of an appropriate field
redefinition. Our work is organized as follows. In section \ref{sec1} we
provide basic definitions and explain our notation. The proof of the
redundancy of the off-shell parameters in chiral EFT, which is the main result
of this work, is presented in section~\ref{sec2}. Some useful relations for
the spin-3/2 differential operator which are needed in the proof are derived
in the appendix.

\section{Basic definitions and notation}
\def\theequation{\arabic{section}.\arabic{equation}}
\setcounter{equation}{0}
\label{sec1}

The free Lagrangian for a massive spin--3/2 field representing the delta isobar
can be written in the form 
\beq
\mathcal{L}_0 = \bar{\psi}(i
    \slsm{\partial}-m)\psi , 
\label{free}
\eeq
where $\psi \equiv \psi_\mu^i$ is a conventional Rarita-Schwinger spinor. 
Here and in what follows, we make use of the
following short-hand notation in order to keep the presentation compact
\beqa
\bar{\psi}O\psi&=&\bar{\psi}^\mu O_{\mu\nu}\psi^\nu=\bar{\psi}^\mu_i
O^{i j}_{\mu\nu}\psi_j^\nu, \nonumber\\
\bar{\psi}G N&=&\bar{\psi}^\mu G_\mu N=\bar{\psi}^\mu_i G_\mu^i N~,
\label{isolorentznotation}
\eeqa
and do not write the isospin and Dirac indices explicitly. The 
derivative operator that appears in the above equation  has the form 
\beq
\label{derivativeDelta}
{\slsm{\partial} }^{i j}_{\mu\nu}=
\gamma_{\mu\nu\alpha}  \partial^\alpha\delta^{i j}
\eeq
while the mass operator reads 
\beq
m^{i j}_{\mu\nu}=\delta^{i j} m \gamma_{\mu \nu}.
\eeq
The $\gamma$-tensors in the above equations are defined according to 
\beqa
\gamma_{\mu\nu} &=&\frac{1}{2}[\gamma_\mu,\gamma_\nu]~,\nonumber\\
\gamma_{\mu\nu\alpha} &=&\frac{1}{2}\{\gamma_{\mu\nu},\gamma_\alpha\}~,
\eeqa
where the $\gamma_\mu$ are the Dirac matrices. Notice that the quantities 
$\gamma_{\mu\nu}$ and $\gamma_{\mu\nu\alpha}$ are completely antisymmetric
with respect to the Dirac indices. 
We emphasize that the free Lagrangian is sometimes written in a more general
form\footnote{For a corresponding expression in $d$ dimensions see
  Ref.~\cite{Pilling:2004cu}.}, 
see e.g.~Ref.~\cite{Moldauer:1955}: 
\beq
\mathcal{L}_0^{(A)} = \bar{\psi}^\mu_i \delta^{ij} 
\Big[ (i \slsm{\partial}  - m ) g_{\mu \nu} + i A (\gamma_\mu \partial_\nu +
\gamma_\nu \partial_\mu ) + \frac{i}{2} ( 3 A^2 + 2 A + 1 ) \gamma_\mu 
 \slsm{\partial} \gamma_\nu + m ( 3 A^2 + 3 A + 1 ) \gamma_\mu \gamma_\nu
 \Big] \psi_j^\nu , 
\label{freeA}
\eeq
where $A\neq -1/2$ is an arbitrary real parameter, see also \cite{Shklyar:2008kt} for a
related recent discussion.  As pointed out in
Ref.~\cite{PascalutsaPhD}, the above Lagrangian can be rewritten in the form
given in Eq.~(\ref{free}) if one redefines the spin-3/2 field as follows
\beq
\psi_i^\mu \to \left( g^{\mu \nu} - \frac{1 + A}{2} \gamma^\mu
  \gamma^\nu \right) \psi_i^\nu\,.
\eeq
We will, therefore, adopt the form given in Eq.~(\ref{free}) in the
following.

\section{Redundancy of the off-shell parameters in chiral EFT}
\def\theequation{\arabic{section}.\arabic{equation}}
\setcounter{equation}{0}
\label{sec2}

The explicit inclusion of the delta isobar into an effective pion-nucleon field
theory is motivated by its strong coupling to the $\pi N \gamma$ system as
well as the small value of the delta-nucleon mass splitting $\Delta = 293
\mbox{ MeV} \simeq 2M_\pi$. Notice, however, that contrary to the pion mass $M_\pi$, the
delta-nucleon mass splitting does not vanish in the chiral
limit. Consequently, such an extended EFT with explicit $\Delta$ isobar degrees of
freedom does not have the same chiral limit as QCD \cite{Gasser:1979hf}. 
One can set up a consistent power counting scheme by treating the 
mass--splitting $\Delta \equiv m_\Delta - m_N$ as an additional small
parameter besides the external
momenta $Q$ and the pion mass $M_\pi$ which is referred to as the small-scale
expansion, see \cite{Hemmert:1997ye} for more details (note that we use the
symbol $\Delta$ for both the delta field and the mass--splitting, but it is
always clear from the context what is meant). Any matrix element or
transition current then possesses  a low-energy expansion of the form 
\beq
\mathcal{M} = \epsilon^n \mathcal{M}_1 + \epsilon^{n+1} \mathcal{M}_2 +
\epsilon^{n+2} \mathcal{M}_3 + \ldots \,,
\eeq
where the power $n$ depends on the process under consideration and $\epsilon$
collects the small parameters 
\beq
\epsilon \in \left\{ \frac{Q}{\Lambda_\chi}, \, \frac{M_\pi}{\Lambda_\chi}, \,
\frac{\Delta}{\Lambda_\chi} \right\}\,.
\eeq
Here, $\Lambda_\chi \sim 1$ GeV refers to the chiral symmetry breaking
scale. 

The most general effective Lagrangian for pions, nucleons and the
delta isobar can be written in the following symbolic form
\beq
\label{lagr}
\mathcal{L}_{\pi \Delta} + \mathcal{L}_{\pi N \Delta} =\bar{\psi}(i
    \sl{D}-m+V)\psi+\bar{\psi}G N+\bar{N}\bar{G} \psi + \ldots, 
\label{sselagrangian}
\eeq
where $N$ is the nucleon field and the ellipses refer to terms which do not
involve the delta isobar and are irrelevant for our work. 
The covariant derivative of the spin-3/2 field has the form 
\beq
\label{derivativeDelta}
\left[i \sl{D}\right]^{i j}_{\mu\nu}=i \, \gamma_{\mu\nu\alpha} (D^{i j})^\alpha
= i\, \gamma_{\mu\nu\alpha} \left[ \partial^\alpha\delta^{i j}+(\Gamma^{i j})^\alpha
\right]
\eeq
where $\Gamma_\alpha^{i j}$ is  the so-called chiral connection whose explicit
form can be found in Ref.~\cite{Hemmert:1997ye}. 
The matrices $V$, $G$ and $\bar G$ in
Eq.~(\ref{lagr}) incorporate all possible local terms allowed by the symmetry
requirements. A finite number of terms contribute at a given order
$\epsilon^n$. Utilizing the formalism by Coleman, Callan, Wess and
Zumino \cite{Coleman:1969sm,Callan:1969sn}, one simply needs to write down all
possible isospin-invariant terms\footnote{Here and in the following, we
  consider the two-flavor case of  up
and down quarks.} constructed from the various building blocks which transform
covariantly under global chiral transformations in order to 
incorporate the non--linearly realized chiral symmetry. 
The explicit form of the
various building blocks which depend on external sources and, in a nonlinear
way, on pion fields as well as their transformation rules under chiral
rotations, parity inversion, hermiticity and charge conjugations are not relevant
for our work and can be found in \cite{Hemmert:1997ye}. 

For the following discussion, we require the pertinent counting rules for the
building blocks of the chiral effective Lagrangian. We employ the scheme of
Ref.~\cite{Hemmert:1997ye}. We emphasize that 
\beq
\label{epsilon0}
m, \; \psi , \; N, \; \gamma_\mu , \; D_\mu \sim \mathcal{O} (\epsilon^0)
\eeq
whereas 
\beq
\label{epsilon1}
i \sl{D}-m \sim \mathcal{O} (\epsilon )\,.
\eeq
and
\beq
\left[D_\mu,D_\nu\right] \sim \mathcal{O} (\epsilon^2 ) \,.
\eeq
Notice further that the interaction terms $V,G$ or 
$\bar{G}$  in 
Eq.~(\ref{sselagrangian}) have a small-scale expansion of the form 
\beq
V=\sum_{n=1}^\infty V^{(n)}\,, \quad \quad 
G=\sum_{n=1}^\infty G^{(n)}\,, \quad \quad 
\bar{G}=\sum_{n=1}^\infty \bar{G}^{(n)}\,, 
\eeq 
with the superscript referring to the power
of $\epsilon$. In the following, we will also frequently use the short-hand
notation for an interaction operator $W$:
\beq
W^{(\le n )} \equiv \sum_{i=1}^n W^{(i)}.
\eeq 

All Rarita-Schwinger fields appearing in the interaction terms of the
effective Lagrangian are accompanied by matrices 
\beq
\theta_{\mu \nu} (z) = g_{\mu \nu} + z \gamma_\mu \gamma_\nu\,,
\eeq
where $z$ denotes an off-shell parameter which governs the coupling of the
(off shell) spin-1/2 components to a given operator. For example, the
interaction $\bar \psi V \psi$ can be written as a series of interactions with
different off-shell parameters  
\beq
\bar{\psi}V\psi=\sum_i \bar{\psi}\Theta(z_i^\prime)V_i\Theta(z_i)\psi + {\rm
  h.c.}.  
\eeq
One can switch to the gauge invariant formulation suggested by Pascalutsa
\cite{Pascalutsa:1998pw} employing the non-linear field redefinition given explicitly in
Ref.~\cite{Krebs:2008zb} 
\beq
\label{redef}
\mathcal{L}=\bar{\psi}(i\slsm{\partial}-m+V)\psi \quad \to 
\quad {\cal L}^\prime=\bar{\psi} \bigg( i\slsm{\partial}-m+i\slsm{\partial}
\frac{1}{m}V\frac{1}{m}\left[1-(i\slsm{\partial}+m)
\frac{1}{m}V\frac{1}{m}\right]^{-1}\hspace{-0.4cm}i\slsm{\partial}\bigg)
\psi\,. 
\eeq
where the inverse mass operator 
in $d$ dimensions has the following form
\beq
\left[\frac{1}{m}\right]^{i j}_{\mu\nu}=-\frac{1}{m}\delta^{i j}
\left(g_{\mu\nu}+\frac{1}{1-d}\gamma_\mu\gamma_\nu\right).
\eeq
Clearly, the last term in the brackets in Eq.~(\ref{redef}) should be
understood in terms of the 
Taylor expansion giving rise to an infinite series of local
interactions. 
The more general case of bilinear and linear couplings  of the spin-3/2
fields can be treated analogously, see Ref.~\cite{Krebs:2008zb}. 
The advantage of the new formulation is that all couplings of the spin-3/2
fields have the same constraints as in the free field theory. In the absence
of the chiral symmetry one would immediately conclude that the dependence of
the off-shell parameters can be completely absorbed into redefinition of
the LECs entering the Lagrangian $\mathcal{L}'$. It is not obvious that this
statement also applies to \emph{chiral} effective Lagrangians since chiral symmetry is not
realized on a term-by-term basis in $\mathcal{L}'$ any more. 

In the following,
we will prove that the off-shell parameters entering the effective
chiral Lagrangian are indeed redundant and can be eliminated by suitably
chosen field redefinitions. To this end we first prove a lemma which
allows to rewrite the interaction terms in the effective Lagrangian at a fixed
order in the  small scale expansion in a more convenient way by making use of the
equations of motion. For doing that, we have to be more specific regarding the
form of the interaction terms entering the effective Lagrangian as compared to the
previous discussion. This will require to introduce further interaction operators
$\bar{V}$ and $W$ in $\mathcal{L}_{\pi \Delta}$ and $H$ and $\bar{H}$ in
$\mathcal{L}_{\pi N \Delta}$ in addition to the ones which have already
been defined in Eq.~(\ref{lagr}). 


\bigskip
\emph{Lemma.} Let $V,\bar{V},W,H,\bar{H},G,\bar{G}$ be 
local interaction operators which appear in the most general chiral effective
Lagrangian for pions, nucleons and the delta isobar with the chiral symmetry
implemented on the term-by-term basis. Then, the following relations are true:
\beqa
{\cal L}_1&=&\bar{\psi}(i \sl{D}-m+V^{(n)}+\bar{V}^{( n)} + W^{(\le n)})\psi + 
\bar{\psi}G^{(\le n)} N + \bar{N}\bar{G}^{(\le n)}\psi+{\cal
  O}(\epsilon^{n+1})
\nn
&\simeq&\bar{\psi}(i \sl{D}-m+i
\sl{D}\frac{1}{m}V^{(n)}+\bar{V}^{(n)}\frac{1}{m}i \sl{D} + W^{(\le n)})\psi + 
\bar{\psi}G^{(\le n)} N + \bar{N}\bar{G}^{(\le n)}\psi+{\cal
  O}(\epsilon^{n+1}),\label{lemma1}\\
{\cal L}_2&=&\bar{\psi}(i \sl{D}-m+ W^{(\le n)})\psi + \bar{\psi} (H^{(n)}+G^{(\le n)}) N
+\bar{N}( \bar{H}^{(n)}+\bar{G}^{(\le n)})\psi+{\cal
  O}(\epsilon^{n+1})\nn
&\simeq&
\bar{\psi}(i \sl{D}-m+ W^{(\le n)})\psi + \bar{\psi} (i \sl{D}\frac{1}{m}H^{(n)}
+G^{(\le n)}) N
+\bar{N}( \bar{H}^{(n)}\frac{1}{m}i \sl{D}+\bar{G}^{(n)})\psi +{\cal
  O}(\epsilon^{n+1}) ,\label{lemma2}
\eeqa
where ${\cal L}_i\simeq{\cal L}_j$ means that the theories based on 
the Lagrangians ${\cal L}_i$ and ${\cal L}_j$ lead to the same 
$S$-matrix (provided the low-energy constants entering $\mathcal{O}
(\epsilon^{n+1})$-terms are re-defied appropriately). 

\bigskip
\emph{Proof.}
The lemma is proved by applying to the left-hand side of Eq.~(\ref{lemma1})
the field transformations 
$$
\psi\rightarrow \psi + \frac{1}{m}V^{(n)}\psi, \quad
\bar{\psi}\rightarrow \bar{\psi}+\bar{\psi}\bar{V}^{(n)}\frac{1}{m},
$$
and to the left-hand side of Eq.~(\ref{lemma2}) the field transformations
\beq
\psi\rightarrow \psi + \frac{1}{m}H^{(n)} N, \quad
\bar{\psi}\rightarrow \bar{\psi}+\bar{ N}\bar{H}^{(n)}\frac{1}{m},
\eeq
where we made use of the counting rules in Eq.~(\ref{epsilon0}). 
These field redefinitions do not change the $S$-matrix by virtue of the
equivalence theorem \cite{Coleman:1969sm,Haag:1958vt}. 

\bigskip
Notice that applying the previous lemma twice one immediately obtains the
following useful relation:
\beqa
&& \bar{\psi}(i \sl{D}-m+V^{(n)}+\bar{V}^{(n)} + W^{(\le n)})\psi + 
\bar{\psi}G^{(\le n)} N + \bar{N}\bar{G}^{(\le n)}\psi+{\cal
  O}(\epsilon^{n+1}) \nn
&\simeq& \bar{\psi}\bigg(i \sl{D}-m+i \sl{D}\frac{1}{m}(V^{(n)}+\bar{V}^{(n)})
\frac{1}{m}i \sl{D}
+ W^{(\le n)}\bigg)\psi 
+
\bar{\psi}G^{(\le n)} N + \bar{N}\bar{G}^{(\le n)}\psi
+ {\cal
  O}(\epsilon^{n+1})\,.
\eeqa
With these preparations we are now in the position to prove the redundancy of the
off-shell parameters in chiral EFT. 

\bigskip
\emph{Theorem.} The interactions $\bar{\psi}V\psi$, $\bar{\psi}W\psi$, $\bar{\psi}X N$, 
$\bar{N} \bar{X}\psi$, $\bar{\psi}Y N$, 
$\bar{N} \bar{Y}\psi$ of the form
\beqa
V^{i j}_{\mu\nu}&=&\gamma_\mu O^{i j}_\nu + \bar{O}^{i j}_\mu \gamma_\nu,
\nonumber \\
X^i_\mu &=& \gamma_\mu Q^i , 
\quad \bar{X}^i_\mu= \bar{Q}^i \gamma_\mu,\nn
W^{i j}_{\mu\nu}&=&D_\mu^{i l} P^{l j}_\nu + \bar{P}^{i l}_\mu D_\nu^{l j},
\nonumber\\
Y^i_\mu &=& D_\mu^{i l} S^l , \quad 
\bar{Y}^i_\mu= \bar{S}^l D_\mu^{l i} ~,
\eeqa
with arbitrary interaction operators $O$, $\bar{O}$, $P$, $\bar{P}$, $Q$,
$\bar{Q}$, $S$, $\bar{S}$ which appear in the most general chiral effective
Lagrangian for pions, nucleons and the delta isobar do not affect $S$-matrix
elements.

\bigskip
\emph{Proof.}
We will prove the theorem by induction making use of the following 
relations for the spin-$3/2$ differential operator which are derived in
appendix:
\beqa
\left[\left( i    \sl{D}\frac{1}{m}\right)^2\right]^{\mu\nu}_{i j}\gamma_\nu&=&
\gamma_\nu\left[\left( \frac{1}{m} i \sl{D}\right)^2\right]^{\nu\mu}_{i j}=
-\frac{1}{2
  m^2}\frac{d-2}{d-1}\gamma^{\mu\alpha\beta}\left[D_\alpha,D_\beta\right]_{i
  j}={\cal O}(\epsilon^2), \nn
\left[\left( i    \sl{D}\frac{1}{m}\right)^2\right]^{\mu\nu}_{i l}D_\nu^{l j}&=&
\frac{1}{2 m}\left[ i    \sl{D}\frac{1}{m}\right]^{\mu\lambda}_{i l}
\gamma_{\lambda \alpha \beta}\left[D^\alpha,D^\beta\right]_{l j}-
\frac{1}{2 m^2}\frac{d-2}{d-1}\gamma^{\mu \alpha \beta}
\left[D_\alpha,D_\beta\right]_{i l}D_\nu^{l j} \gamma^\nu 
= {\cal O}(\epsilon^2),\nn
D_\nu^{i l}\left[\left( \frac{1}{m} i \sl{D}\right)^2\right]^{\nu\mu}_{l j}&=&
\frac{1}{2 m}\gamma_{\lambda \alpha \beta}\left[D^\alpha,D^\beta\right]_{i l}
\left[ \frac{1}{m} i \sl{D}\right]^{\lambda\mu}_{l j}
-\frac{1}{2 m^2}\frac{d-2}{d-1}D_\nu^{i l} \gamma^\nu
\gamma^{\mu \alpha \beta}
\left[D_\alpha,D_\beta\right]_{l j}
={\cal O}(\epsilon^2).\label{diffop3half}
\eeqa
Let us now show by induction that the interaction 
\beq
Z_{\mu\nu}^{i j}=V_{\mu \nu}^{i j}+W_{\mu \nu}^{i j}.
\eeq
in the bilinear couplings of the delta isobar can be transformed away from the
effective Lagrangian via an appropriately chosen field redefinition. 

\bigskip
\emph{Induction start.} The most general bilinear interaction terms at first
order in the small scale expansion (apart from the interactions entering the
covariant derivative in the kinetic term) can be written in the form 
$\bar \psi (Z^{(1)}+R^{(1)} ) \psi$, where the operator $R^{(1)}$ does not involve the
structures present in $Z^{(1)}$, i.~e.~does not contain contributions which
vanish for on-shell spin-3/2 fields and is, therefore, free of the off-shell parameters. 
Applying the previous lemma several times we get
\beqa
{\cal L}&=&\bar{\psi}(i \sl{D} - m + Z^{(1)}+R^{(1)})\psi + \bar{\psi}G^{(1)}N
+\bar{N}\bar{G}^{(1)}\psi+{\cal O}(\epsilon^2)\nn
&\simeq&\bar{\psi}(i \sl{D} - m + \left( i    \sl{D}\frac{1}{m}\right)^2 Z^{(1)}
\left( \frac{1}{m} i
    \sl{D}\right)^2+R^{(1)})\psi + \bar{\psi}G^{(1)}N
+\bar{N}\bar{G}^{(1)}\psi+{\cal O}(\epsilon^2)\nn
&=&\bar{\psi}(i \sl{D} - m +R^{(1)})\psi + \bar{\psi}G^{(1)}N
+\bar{N}\bar{G}^{(1)}\psi+{\cal O}(\epsilon^2)
\eeqa
In the last step we used 
$$
\left( i    \sl{D}\frac{1}{m}\right)^2 Z^{(1)}
\left( \frac{1}{m} i
    \sl{D}\right)^2={\cal O}(\epsilon^3).
$$

\bigskip
\emph{Induction assumption.} 
We assume that the operator $R^{(\le n)}$ does not include structures present
in $Z^{(\le n)}$ and that the equality
\beqa
{\cal L}&=&\bar{\psi}(i \sl{D} - m + Z^{(n)}+R^{(\le n)})\psi + \bar{\psi}G^{(\le
  n)}N
+\bar{N}\bar{G}^{(\le n)}\psi+{\cal O}(\epsilon^{n+1})\nn
&\simeq&\bar{\psi}(i \sl{D} - m +R^{(\le n)})\psi + \bar{\psi}G^{(\le
  n)}N
+\bar{N}\bar{G}^{(\le n)}\psi+{\cal O}(\epsilon^{n+1})
\eeqa
 holds true.

\bigskip
\emph{Induction step.} 
By the induction assumption, we can write the Lagrangian in
the form 
\beqa
{\cal L}&\simeq&\bar{\psi}(i \sl{D} - m + Z^{(n+1)}+R^{(\le n+1)})\psi + \bar{\psi}G^{(\le
  n+1)}N
+\bar{N}\bar{G}^{(\le n+1)}\psi+{\cal O}(\epsilon^{n+2})\nonumber
\eeqa
where the operator $R^{(\le n+1)}$ does not include structures of the 
form $Z^{(\le n+1)}$.
Applying again the previous lemma several times we
get
\beqa
{\cal L}&\simeq&\bar{\psi}(i \sl{D} - m + Z^{(n+1)} 
+R^{(\le n+1)})\psi + \bar{\psi}G^{(\le
  n+1)}N
+\bar{N}\bar{G}^{(\le n+1)}\psi+{\cal O}(\epsilon^{n+2})\nn
&\simeq&\bar{\psi}(i \sl{D} - m + \left( i    \sl{D}\frac{1}{m}\right)^2 Z^{(n+1)}\left( \frac{1}{m} i
    \sl{D}\right)^2+R^{(\le n+1)})\psi 
+ \bar{\psi}G^{(\le n+1)}N
+\bar{N}\bar{G}^{(\le n+1)}\psi+{\cal O}(\epsilon^{n+2})\nn
&=&\bar{\psi}(i \sl{D} - m +R^{(\le n+1)})\psi 
+ \bar{\psi}G^{(\le n+1)}N
+\bar{N}\bar{G}^{(\le n+1)}\psi+{\cal O}(\epsilon^{n+2}) \,,
\eeqa
where in the last step we used the relation
\beq
\left( i    \sl{D}\frac{1}{m}\right)^2 Z^{(n+1)}
\left( \frac{1}{m} i
    \sl{D}\right)^2={\cal O}(\epsilon^{n+3})\, .
\eeq

\bigskip
In the same way one can show that the interactions $X,\bar{X},Y,\bar{Y}$ 
can be transformed away which completes the proof of the theorem.

\bigskip
As an application, consider the (leading) dimension-one effective Lagrangian
describing the coupling of the massive spin-3/2 fields to the pions which has
the form \cite{Hemmert:1997ye}
\beq
\label{lagr1}
\mathcal{L}_{\pi \Delta}^{(1)} = \bar \psi^\mu_i  \left(  
\left[i \sl{D}\right]^{i j}_{\mu\nu} - m_{\mu \nu}^{ij}  
- \frac{g_1}{2} g_{\mu \nu} \slsm{u}^{ij} \gamma_5 - 
\frac{g_2}{2} \left( \gamma_\mu u_\nu^{ij} + u_\mu^{ij} \gamma_\nu \right) -
\frac{g_3}{2} \gamma_\mu  \slsm{u}^{ij} \gamma_5 \gamma_\nu \right)
\psi^\nu_j\,,
\eeq
where the $g_i$ are LECs, $u_\mu^{ij} = u_\mu \delta^{ij}$, $u_\mu = i
(u^\dagger \partial_\mu u - 
i \partial_\mu u^\dagger )$ and the matrix $U(x) = u^2 (x)$ collects the pion
fields, see \cite{Hemmert:1997ye} for more details. Notice that $g_2$ and
$g_3$ describe pion couplings to off-mass-shell components of the spin-3/2
fields. It immediately follows from our analysis that both the $g_2$- and $g_3$-terms are
redundant, i.e.~they do not contribute to S-matrix elements calculated from
Eq.~(\ref{lagr1}) (and higher-order counter terms) using naive Feynman rules.

\section*{Acknowledgments}

The work of E.E. and H.K. was supported in parts by funds provided from the 
Helmholtz Association to the young investigator group  
``Few-Nucleon Systems in Chiral Effective Field Theory'' (grant  VH-NG-222)
and through the virtual institute ``Spin and strong QCD'' (grant VH-VI-231). 
This work was further supported by the DFG (SFB/TR 16 ``Subnuclear Structure
of Matter'') and by the  EU-Research Infrastructure Integrating Activity
 ``Study of Strongly Interacting Matter'' (HadronPhysics2, grant n. 227431)
under the Seventh Framework Program of the EU.




\begin{appendix}
\section{Properties of spin-$3/2$ differential operator}
\def\theequation{A.\arabic{equation}}
\setcounter{equation}{0}
\label{diffopapp}

In this appendix we derive the relations for the spin-$3/2$ differential
operator in Eq.~(\ref{diffop3half}). We first contract the
inverse of the mass operator with the Dirac-matrix $\gamma^\mu$ leading to:
\beq
\gamma_\nu\left[\frac{1}{m}\right]^{\nu\alpha}_{i j}=-\frac{1}{m}\gamma_\nu
\left(g^{\nu\alpha}+\frac{1}{1-d}\gamma^\nu\gamma^\alpha\right)\delta_{i j}=
-\frac{1}{m}\, \frac{1}{1-d}\gamma^\alpha\delta_{i j}.
\eeq
Multiplying this result with the operator $i \sl{D}$ we obtain
\beq
\gamma_\nu\left[\frac{1}{m}\right]^{\nu\alpha}_{i k}
\left[i \sl{D}\right]^{k j}_{\alpha\beta}=-\frac{1}{m}\frac{1}{1-d}\gamma^\alpha
\gamma_{\alpha\beta\lambda} i D^\lambda_{i j}=-\frac{1}{m}\frac{d-2}{1-d}
\gamma_{\beta\lambda} i D^\lambda_{i j}=\frac{1}{m^2}\frac{d-2}{1-d} 
i D^\lambda_{i k}
m_{\lambda\beta}^{k j}.\label{idoverm}
\eeq
In the second step of the last equation we used the identity
\beq
\gamma^\alpha\gamma_{\alpha\beta\lambda}=(d-2)\gamma_{\beta\lambda}.
\eeq
Multiplying Eq.~(\ref{idoverm}) with the inverse of the mass operator from
the right we get
\beq
\gamma_\nu\left[\frac{1}{m}i \sl{D}\frac{1}{m}\right]^{\nu\alpha}_{i j}=
\frac{1}{m^2}\frac{d-2}{1-d} i D^\alpha_{i j}.
\eeq
Finally, multiplication of the last equation with the operator $i \sl{D}$
from the right yields
\beq
\gamma_\nu\left[\left(\frac{1}{m}i \sl{D}\right)^2\right]^{\nu\mu}_{i j}=
\frac{1}{m^2}\frac{d-2}{1-d} i D_\alpha^{i k}\gamma^{\alpha\mu\beta} i
D_\beta^{k j}=
-\frac{1}{2 m^2}\frac{d-2}{d-1}\gamma^{\mu\alpha\beta}[D_\alpha,D_\beta]_{i j}.
\eeq
where the commutator of the differential operators is given by
\beq
[D_\alpha,D_\beta]_{i j}=D_\alpha^{i k}D_\beta^{k j}-D_\beta^{i k} D_\alpha^{k j}.
\eeq
All other relations given in Eq.~(\ref{diffop3half}) can be derived in a
completely analogous way. 
\end{appendix}


\end{document}